\documentclass[english]{JHEP3}
\usepackage[T1]{fontenc}
\usepackage[latin1]{inputenc}
\usepackage{graphicx}
\usepackage{amssymb}
\usepackage{babel}
\usepackage{cite}
\usepackage{epsfig}


\title{The Friedman Universe with the stochastic cosmological constant}

\author{R. Manka\\ University of Silesia, Institute of Physics, Katowice, Poland\\ E-mail:manka@us.edu.pl}
\author{D. Panchyrz\\ University of Silesia, Institute of Physics, Katowice, Poland}

\abstract{
The Friedman Universe with the remnant stochastic scalar field in
the hybrid inflation model is examined. It is shown that the small
effective cosmological constant~$\delta\Omega_{\Lambda}$~appears
which increases the cosmological expansion.
}

\keywords{Classical Theories of Gravity, Stochastic Processes}

\begin{document}

\section{Introduction}

The recent astrophysical observation of the accelerating expansion
\cite{perl,reiss} of the Universe indicates that the cosmological
constant $\Lambda$ has a small nonzero value. However, the origin
of the cosmological constant $\Lambda$ is still unclear \cite{weinberg}.
Its origin may be connected, for example, with the presence of some  scalar fields \cite{sami}.
The string inspired gravity is much richer than   the Einstein one including
several types of scalar fields (dilaton or moduli fields). 
What is more, the interaction
with the background strings may  also a produce chaos \cite{Vilenkin}.
Similarly, the cosmological constant may gain small stochastic component \cite{musso}. 

The main aim of this work is to examine how small stochastic fluctuation
influence the cosmological expansion.

\section{The classical cosmology}

Let us start with the general gravity with some scalar field $\phi$
and  ordinary dustlike matter. The scalar field $\phi$ may be the
remnant of the hybrid inflation model \cite{liddle} (the {}``waterfall''
field). In the original hybrid inflation model \cite{liddle} we have
two scalar fields: the inflation field $\varphi$ and the {}``waterfall''
field $\phi$ described by the potential 
\begin{equation}
U(\varphi,\phi)=\frac{1}{2}\lambda\left(\phi^{2}-\phi_{0}^{2}\right)+\frac{1}{2}m^{2}\varphi^{2}+\frac{1}{2}\lambda'\,\phi^{2}\varphi^{2}.
\end{equation}
Fluctuations of the  $\varphi$ late in the inflation period deminishes and
$\phi$ goes to $\phi_{0}$. However, they  may be a source
of the chaotic fluctuation  at the present time. The Lagrange function of the system 
has correspondingly three terms
\begin{eqnarray}
 & L=L_{g}+L_{\phi}+L_{m}\label{eq:lag1}\\
 & L_{g}=-\frac{1}{2\kappa}R\label{eq:lag2}\\
 & L_{\phi}=\frac{1}{2}\partial_{\mu}\phi\partial^{\mu}\phi-U(\phi)\label{eq:lag3}
\end{eqnarray}
which describes the Einstein gravity $L_g$, the {}``waterfall''
field $L_{\phi}$ and the dustlike matter $L_m$ 
The scalar potential \begin{equation}
U(\phi)=U_{0}(\phi)+\epsilon J_{s}\phi\label{eq:upot}\end{equation}
consists of the nonlinear part $U_{0}(\phi)$ with a local minimum
at $\phi_{0}$ and a scalar external source $J_{s}$ with a scalar
charge $\epsilon$. The scalar potential $U_{0}(\phi)$ is  the remnant
from the inflation epoch. The main assumption of this work is that
the scalar external source $J_{s}$ has the stochastic nature. 

The Euler-Lagrange equations of the system (\ref{eq:lag1}) give the
Einstein's equation
\begin{equation}
R_{\mu\nu}-{\frac{1}{2}}Rg_{\mu\nu}=-\kappa T_{\mu\nu}=-\kappa\left(T_{\mu\nu}^{m}+T_{\mu\nu}^{\phi}\right)\label{einstein}
\end{equation}
where $\kappa=8\pi G$ (in $c=\hbar=1$ units) together with the equation
for the scalar field $\phi$. In this equation the fluid is described
by the energy-momentum tensor in the standard form $T_{\mu\nu}^{m}=(\rho+p)u_{\mu}u_{\nu}+Pg_{\mu\nu}$
( $u^{\mu}$ is the fluid four-velocity, $\rho$ is the energy density
in the rest frame of the fluid and $P$ is the pressure in that same
frame). For scalar field the energy-momentum tensor has the form $T_{\mu\nu}^{\phi}=\partial_{\mu}\phi\partial_{\nu}\phi-g_{\mu\nu}L_{\phi}$. 

The Lagrange-Euler equation of the $\phi$ field is 
\begin{equation}
\square\phi=-\frac{\partial L_{\phi}}{\partial\phi}=\frac{\partial U_{0}}{\partial\phi}+\epsilon J_{s}\label{eq:scal1}
\end{equation}
where  $\square f=-\frac{1}{\sqrt{-g}}\partial_{\mu}(\sqrt{-g}g^{\mu\nu}\partial_{\nu}f)$
is the d'Alembert operator in a curved space-time. In vacuum (when $J_{s}=0)$ the static local
extremum $\phi_{0}$ generates the effective cosmological constant
$\Lambda^{0}$ with $\Omega_{\Lambda}^{0}=\rho_{\phi}^{0}/\rho_{c}$,
with $\rho_{\phi}^{0}=U(\phi_{0})$. 

The space-time interval $ds$ of the Friedman-Robertson-Wolker (FRW)
metric is written in the standard way 
\begin{equation}
ds^{2}=g_{\mu\nu}dx^{\mu}dx^{\nu}=c^{2}dt^{2}-a^{2}(t)\left[\frac{dr^{2}}{1-kr^{2}}+r^{2}\left(d\theta^{2}+\sin^{2}\theta d\phi^{2}\right)\right]\label{frwmetric2}
\end{equation}
Calculating the Einstein's equations (\ref{einstein}) for the metric
(\ref{frwmetric2}) we get two independent equations. The first is
known as the Friedman equation, \begin{equation}
H^{2}\equiv\left(\frac{{\dot{a}}}{a}\right)^{2}=\frac{8\pi G}{3}\rho_{m}+\frac{\Lambda}{3}-\frac{kc^{2}}{a^{2}}=\frac{8\pi G}{3}\left(\rho_{m}+\rho_{\Lambda}\right)-\frac{kc^{2}}{a^{2}}\label{Friedmann}\end{equation}
Here the over-dot denotes a derivative with respect to cosmic time
$t$ . This equation is a constraint equation, in the sense that we
are not allowed to freely specify the time derivative $\dot{a}$;
it is determined in terms of the energy density and curvature. The
second equation, which is an evolution equation, is often combined
with the first one (\ref{Friedmann}) to obtain the \textit{acceleration
equation} \begin{equation}
\frac{{\ddot{a}}}{a}=-\frac{4\pi G}{3}\left(\rho+3\frac{P}{c^{2}}\right)+\frac{\Lambda}{3}\label{acceleration}\end{equation}
The equation (\ref{eq:scal1}) for the scalar field $\phi$ now takes
the form\begin{equation}
\ddot{\phi}+3\frac{\dot{a}}{a}\dot{\phi}=\frac{\partial U_{0}}{\partial\phi}+\epsilon J_{s}\label{eq:scal2}\end{equation}
 The substitution \begin{equation}
a(t)=exp(\sigma(\tau))\label{eq:scal}\end{equation}
transforms the Friedman equation (\ref{Friedmann})  into the following
equation\begin{equation}
\left(\frac{d\sigma}{d\tau}\right)^{2}=\Omega_{m}e^{-3\sigma}+\Omega_{\Lambda}\label{eq:fried}\end{equation}
where $t=t_{H}\tau=H_{0}^{-1}\tau$ and $H_{0}$ is the Hubble constant
at the present time $t_{0}$ and $\Omega=\rho/\rho_{c}$ is the density
parameter (in case of  the flat universe ($k=0$) in dust era ($P=0$)).
The acceleration equation (\ref{acceleration}) together
with the first Friedman equation (\ref{eq:fried}) give
\begin{equation}
\frac{d^{2}\sigma}{d\tau^{2}}=-\frac{3}{2}\Omega_{m}e^{-3\sigma}\label{eq:fried2}
\end{equation}
Close to the minimum the potential $U_{0}(\phi)\sim\frac{1}{2}m_{s}^{2}(\phi-\phi_{0})^{2}$
with $m_{s}^{2}=2\lambda\phi_{0}^{2}.$ Equation for the scalar field
(\ref{eq:scal1}) now takes the form 
\begin{equation}
\frac{d^{2}\phi}{d\tau^{2}}+3\frac{d\sigma}{d\tau}\frac{d\phi}{d\tau}=-\frac{1}{H_{0}^{2}}\frac{\partial U_{0}}{\partial\phi}-\epsilon\eta\sim-\mu^{2}(\phi-\phi_{0})-\phi_{0}\eta\label{lagandre}
\end{equation}
where 
\begin{equation}
\mu=\frac{m_{s}}{H_{0}},\qquad\eta=\frac{J{}_{s}}{\phi_{0}H_{0}^{2}}\label{eta}
\end{equation}
is the dimensionless scalar mass and source term. The main assumption of this
work is that the scalar field close to the local extremum has partly
small stochastic contribution coming from the stochastic source $\eta$.
In the result the effective cosmological term with a small stochastic
contribution originated from the stochastic part of the scalar density
$\delta\rho=\epsilon J_{s}\phi_{0}$ \begin{equation}
\Omega_{\Lambda}=\Omega_{\Lambda}^{0}+\epsilon\lambda\eta\label{eq:decomp}\end{equation}
appears with the dimensionless factor \begin{equation}
\lambda=\frac{8\pi}{3}\left(\frac{\phi_{0}}{M_{Pl}}\right)^{2}\label{lambda}\end{equation}
In cosmology the equation (\ref{lagandre}) plays the role of the Lagrange equation. The stochastic fluctuations of the cosmological constant
will produce the change of scalar field\begin{equation}
\phi=\phi_{0}+\epsilon\phi_{0}F(\tau)\label{eq:phi}\end{equation}
and metric
\begin{equation}
\sigma\left(t\right)=\sigma_{0}\left(t\right)+\epsilon\left(\frac{d\sigma_{0}(\tau)}{d\tau}\right)X\left(\tau\right)\label{eq:decomp2}.\end{equation}
Here $\epsilon$ is a small perturbation parameter ($\epsilon\ll1$). The
fluctuation of the $\sigma(t)$ field around the classical $\sigma_{0}(t)$
means that also the scale factor will fluctuate around the classical
one ($a_{0}(t)=exp(\sigma_{0}(\tau))$) \[
a\left(t\right)=a_{0}\left(t\right)(1+\epsilon X\left(t\right))\]
together with metric\[
g_{\mu\nu}=g_{\mu\nu}^{0}+\epsilon h_{\mu\nu}\]
where \[
h_{\mu\nu}=\left(\begin{array}{cccc}
0\\
 & -2a_{0}(t)X(t)\\
 &  & -2a_{0}(t)X(t)\\
 &  &  & -2a_{0}(t)X(t)\end{array}\right)\]
describes the space fluctuation around the classical metric \[
g_{\mu\nu}^{0}=\left(\begin{array}{cccc}
1\\
 & -a_{0}^{2}(t)\\
 &  & -a_{0}^{2}(t)\\
 &  &  & -a_{0}^{2}(t)\end{array}\right)\]
The linearization of the equation for the scalar field (\ref{lagandre})
gives 
\begin{equation}
 \frac{d^{2}F}{d\tau^{2}}+3\frac{d\sigma_{0}(\tau)}{d\tau}\frac{dF}{d\tau}+\mu^{2}F=-\eta
\label{etaa}
\end{equation}
The linearization (eqs. \ref{eq:decomp} and \ref{eq:decomp2}) using
the equation (\ref{eq:fried})  leads to the
stochastic differential equation 
\begin{equation}
\frac{dX}{d\tau}=\frac{\lambda}{2\left(\frac{d\sigma(\tau)}{d\tau}\right)^{2}}\eta\label{eq:stoch}
\end{equation}
When $\eta$ represents the white noise ($\delta\Omega_{\Lambda}=\eta$) its correlation function obeys 
\begin{equation}
<\eta(\tau)\eta(\tau')>=D\delta(\tau-\tau')
\label{etad}
\end{equation}
with $\delta(\tau)$ being the Dirac distribution.

 The solution of
the deterministic differential equation
\begin{equation}
\left(\frac{d\sigma_{0}}{d\tau}\right)^{2}=\Omega_{m}e^{-3\sigma_{0}(\tau)}+\Omega_{\Lambda}^{0}\label{eq:deter}
\end{equation}
with the present time $t_0$ ( or $\tau_0$) defined as ($a_{0}(\tau_{0})=0$ or
$a_{0}(t_{0})=1$)\begin{equation}
\tau_{0}=\frac{t_{0}}{t_{H}}=H_{0}t_{0}=\frac{2}{3\sqrt{\Omega_{\Lambda}^{0}}}arcsinh(\sqrt{\omega_{0}})\label{eq:age}\end{equation}
and $\omega_{0}=\Omega_{\Lambda}^{0}/\Omega_{m}$ defined by the
deterministic cosmological constant $\Omega_{\Lambda}^{0}=<\Omega_{\Lambda}>$.
The deterministic solution is \begin{equation}
\sigma_{0}(\tau)=\frac{2}{3}\ln\left(\frac{1}{\sqrt{\omega_{0}}}sinh\left(\frac{3}{2}\sqrt{\Omega_{\Lambda}}\tau\right)\right)\label{sigmao}\end{equation}
 and\begin{equation}
a_{0}(\tau)=\left(\frac{sinh(\frac{3}{2}\sqrt{\Omega_{\Lambda}}\tau)}{\sqrt{\omega_{0}}}\right)^{\frac{2}{3}}\label{ao}\end{equation}

\section{The stochastic solution }

The linear stochastic differential equation (SDE) \cite{gard} (\ref{eq:stoch})
may be rewritten in the following form\begin{equation}
dX\left(\tau\right)=f(\tau)X\left(\tau\right)d\tau+d(\tau)dW_{\tau}\label{eq:diffstoch}\end{equation}
where $f(\sigma_{0}(\tau)$ is the drift coefficient , $d(\tau)$
is the diffusion coefficient and $W_{\tau}$ is a Wiener process which
represents intrinsic noise (white noise) ($\delta\Omega_{\Lambda}dt\rightarrow dW_{t}$).
These differential equations are easy to solve numerically \cite{maple}.
The formal solution of the equation (\ref{eq:diffstoch}) has the following
form
\begin{equation}
X(\tau)=\exp{\left(\int_{\tau_{0}}^{\tau}ds\, f\left(s\right)\right)}\left(X\left(\tau_{0}\right)+\int_{\tau_{0}}^{\tau}dW_{s}\exp{\left(-\int_{\tau_{0}}^{\tau}ds\, f\left(s\right)\right)}d(s)\right)\label{eq:solus}
\end{equation}
The linearized  equation (\ref{eq:stoch}) is the generalized  stochastic equation (\ref{eq:diffstoch}) with:
\begin{eqnarray}
 &\nonumber f(\tau)=0\\
 & d(\tau)=\frac{\lambda}{2(\frac{d\sigma_{0}(\tau)}{d\tau})^{2}}
 \end{eqnarray}
where the diffusion coefficient depends straightforward on $\sigma_{0}(\tau)$ (\ref{sigmao}).
 The solution for $X\left(\tau\right)$ takes the simple form:
\[
X\left(\tau\right)=X\left(\tau_{0}\right)+\frac{\lambda}{2}\int_{\tau_{0}}^{\tau}\frac{1}{\left(\frac{d\sigma_{0}(s)}{ds}\right)^{2}}dW_{s}\]
The mean value of $X\left(\tau\right)$ takes the form:
\[
\langle X\left(\tau\right)\rangle=X\left(\tau_{0}\right)\]
According to the equation (\ref{eq:decomp2}) the perturbative solution for $\sigma(\tau)$
may be rewritten as \begin{eqnarray*}
 & \sigma\left(\tau\right)=\sigma_{0}\left(\tau\right)+\epsilon\frac{d\sigma_{0}(\tau)}{d\tau}\left(X\left(\tau_{0}\right)+\frac{\lambda}{2}\int_{\tau_{0}}^{\tau}\frac{1}{\left(\frac{d\sigma_{0}(s)}{ds}\right)^{2}}dW_{s}\right)\\
 & =\sigma_{0}\left(\tau\right)+\epsilon\frac{H_{0}\left(\tau\right)}{H_{0}}\left(X\left(\tau_{0}\right)+\frac{\lambda}{2}\int_{\tau_{0}}^{\tau}\frac{1}{\left(\frac{d\sigma_{0}(s)}{ds}\right)^{2}}dW_{s}\right)\end{eqnarray*}
where $a_{0}(\tau)=exp(\sigma_{0}(\tau))$ is the deterministic scale
factor and $H_{0}(\tau)=\dot{a}_{0}(\tau)/a_{0}(\tau)$ is the deterministic
Hubble parameter. 
\[
\frac{d\sigma_{0}(\tau)}{d\tau}=\sqrt{\Omega_{\Lambda}}\frac{1}{tanh\left(\frac{3}{2}\sqrt{\Omega_{\Lambda}}\tau\right)}\]
When the stochastic fluctuations are small enought ($\delta a(\tau)$)
we get \begin{equation}
a(\tau)=a_{0}(\tau)+\delta a(\tau)=a_{0}(\tau)+\epsilon a_{0}(\tau)\frac{H_{0}\left(\tau\right)}{H_{0}}\left(X\left(\tau_{0}\right)+\frac{\lambda}{2}\int_{\tau_{0}}^{\tau}\frac{1}{\left(\frac{d\sigma_{0}(s)}{ds}\right)^{2}}dW_{s}\right)+...\label{eq:scalefac}\end{equation}
The average calculated with respect to the stochastic process gives
\begin{eqnarray*}
<a(\tau)>=a_{0}(\tau)(e^{\epsilon X_{0}}+\\
+\frac{1}{4}(\epsilon\lambda)^{2}\left(\frac{H_{0}\left(\tau\right)}{H_{0}}\right)^{2}\,&\int_{\tau_{0}}^{\tau}\frac{ds}{\left(\frac{d\sigma_{0}(s)}{ds}\right)^{2}}\int_{\tau_{0}}^{\tau}\frac{dz}{\left(\frac{d\sigma_{0}(z)}{dz}\right)^{2}}<\eta(s)\eta(z)>+...
\end{eqnarray*}
Using (\ref{etad}) and assuming that $X_{0}=X(\tau_{0})=0$
we get
\begin{equation}
<a(\tau)>=a_{0}(\tau)\left(1+\frac{1}{4}(\epsilon\lambda)^{2}D\left(\frac{H_{0}\left(\tau\right)}{H_{0}}\right)^{2}\int_{\tau_{0}}^{\tau}\frac{ds}{\left(\frac{d\sigma_{0}(s)}{ds}\right)^{4}}+...\right)
\label{figdet}
\end{equation}
For late evolutionary times 
\[
<a(\tau)>=a_{0}(\tau)\left(1+\frac{1}{4}(\epsilon\lambda)^{2}D\frac{1}{\Omega_{\Lambda}^{2}}\left(\frac{H_{0}\left(\tau\right)}{H_{0}}\right)^{2}(\tau-\tau_{0})+...\right)\]
 So, if the cosmological constant $\Lambda$ has a stochastic component
$\delta\Omega_{\Lambda}$ then the average scale factor $<a(\tau)>$
will grow more strongly than in the deterministic case (Fig. \ref{fig:Fig1}).%

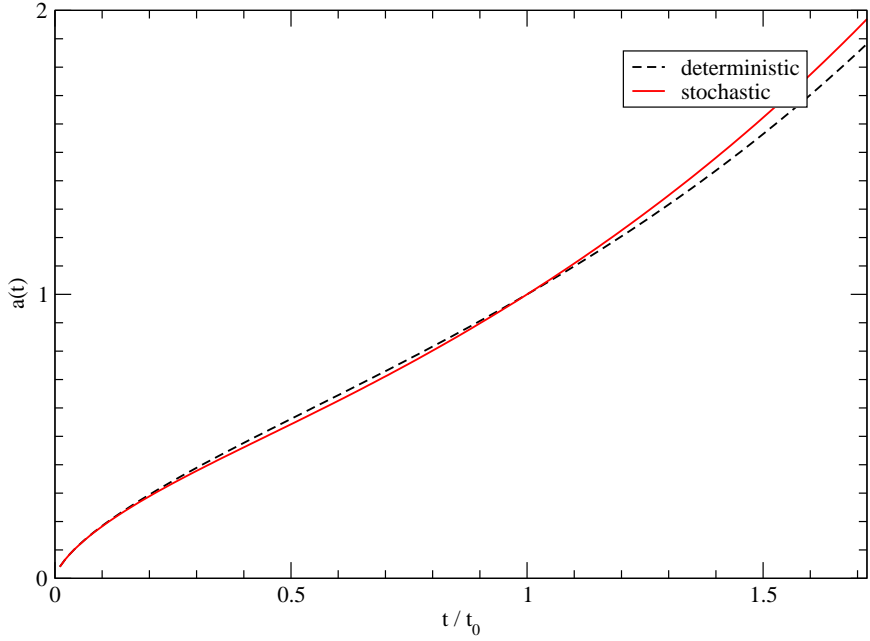
\FIGURE[ht]{\epsfig{file=wyk.eps,scale=0.5}\caption{The scale factor variation with time for deterministic 
(the black curve) and stochastic expansions (the red one (\ref{figdet})). ($X(\tau_0)=0$, $\epsilon=0.5$, $D=5.0$, $\lambda=0.5$ for example) \label{fig:Fig1}}
}

\section{Conclusion}

In conclusion, the {}``waterfall'' field in the hybrid inflation
model as a source of the stochastic process will influence and increase
the Universe expansion. It is also a source of the small effective
cosmological constant $\Lambda$ (\ref{eq:decomp}).

\end{document}